\documentclass{emulateapj}

\pdfoutput=1
\usepackage{epstopdf}
\usepackage{amsmath}

\usepackage{color}

\slugcomment{\today}

\newcommand{\lta}{\lesssim}

\def\arcdeg{\hbox{$^\circ$}}

\def\arcsec{\hbox{$^{\prime\prime}$}}

\def\snid{\ifmmode{\rm \tt SNID}\else{\tt SNID}\fi}
\def\dm15{\ifmmode{\Delta m_{15}}\else{$\Delta m_{15}$}\fi}
\def\magarcsec2{\ \rm{mag \ arcsec}^{-2}}
\newcommand{\about}{$\sim\!\!$~}
\newcommand{\kms}{\,km\,s$^{-1}$}

\shorttitle{Cas~A Was an Asymmetric Supernova}
\shortauthors{Rest et al.}

\begin{document}

\title{Direct Confirmation of the Asymmetry of the Cas~A Supernova with Light Echoes}

\author{A. Rest\altaffilmark{1,2},
R.~J. Foley\altaffilmark{3,4},
B. Sinnott\altaffilmark{5}, 
D.~L. Welch\altaffilmark{5}, 
C. Badenes\altaffilmark{6,7}, 
A.~V. Filippenko\altaffilmark{8}, 
M. Bergmann,
W.~A. Bhatti\altaffilmark{9},
S. Blondin\altaffilmark{10},
P. Challis\altaffilmark{11},
G. Damke\altaffilmark{11},
H. Finley\altaffilmark{12},
M.~E. Huber\altaffilmark{9},
D. Kasen\altaffilmark{13,14},
R.~P. Kirshner\altaffilmark{3},
T. Matheson\altaffilmark{15},
P. Mazzali\altaffilmark{16,17,18}, 
D. Minniti\altaffilmark{19},
R. Nakajima\altaffilmark{20},
G. Narayan\altaffilmark{1},
K. Olsen\altaffilmark{15},
D. Sauer\altaffilmark{21},
R.~C. Smith\altaffilmark{15,22}, and
N.~B. Suntzeff\altaffilmark{23}
}



\altaffiltext{1}{Department of Physics, Harvard University, 17 Oxford
  Street, Cambridge, MA 02138, USA}

\altaffiltext{2}{Space Telescope Science Institute, 3700 San Martin
Dr., Baltimore, MD 21218, USA}

\altaffiltext{3}{Harvard-Smithsonian Center for Astrophysics, 60
  Garden Street, Cambridge, MA 02138, USA}

\altaffiltext{4}{Clay Fellow}

\altaffiltext{5}{Department of Physics and Astronomy, McMaster University,
  Hamilton, Ontario, L8S 4M1, Canada}

\altaffiltext{6}{Benoziyo Center for Astrophysics, Faculty of Physics,
  Weizmann Institute of Science, 76100 Rehovot, Israel}

\altaffiltext{7}{School of Physics and Astronomy, Tel-Aviv University,
  69978 Tel-Aviv, Israel}

\altaffiltext{8}{Department of Astronomy, University of California,
  Berkeley, CA 94720-3411, USA}

\altaffiltext{9}{Department of Physics and Astronomy, Johns Hopkins
  University, Baltimore, 3400 North Charles Street, MD 21218, USA}

\altaffiltext{10}{Centre de Physique des Particules de Marseille (CPPM),
  Aix-Marseille Universit\'e, CNRS/IN2P3, 163 avenue de Luminy, 13288
  Marseille Cedex 9, France}

\altaffiltext{11}{Department of Astronomy, University of Virginia,
  Charlottesville, VA 22904-4325, USA}

\altaffiltext{12}{Department of Physics, Drexel University, 3141
  Chestnut Street, Philadelphia, PA 19104, USA}

\altaffiltext{13}{UCO/Lick Observatory, University of California, Santa
  Cruz, 1156 High Street, Santa Cruz, CA 95064, USA}

\altaffiltext{14}{Hubble Fellow}

\altaffiltext{15}{National Optical Astronomy Observatory, 950 North Cherry
  Avenue, Tucson, AZ 85719-4933, USA}

\altaffiltext{16}{Max-Planck-Institut f\"{u}r Astrophysik,
  Karl-Schwarzschild-Stra\ss e 1, 85741 Garching, Germany}

\altaffiltext{17}{Scuola Normale Superiore, Piazza Cavalieri 7, 56127
  Pisa, Italy}

\altaffiltext{18}{INAF -- Oss. Astron. Padova, vicolo dell'Osservatorio
  5, 35122 Padova, Italy}

\altaffiltext{19}{Vatican Observatory, V00120 Vatican City State,
  Italy}

\altaffiltext{20}{Space Sciences Laboratory, University of California,
  Berkeley, CA 94720-7450, USA}

\altaffiltext{21}{Department of Astronomy, Stockholm University,
  AlbaNova University Center, SE-106 91 Stockholm, Sweden}

\altaffiltext{22}{Cerro Tololo Inter-American Observatory, National
  Optical Astronomy Observatory, Colina el Pino S/N, La Serena,
  Chile}

\altaffiltext{23}{Department of Physics, Texas A\&M University, College
  Station, TX 77843-4242, USA}

 
\begin{abstract}
We report the first detection of asymmetry in a supernova (SN)
photosphere based on SN light echo (LE) spectra of Cas~A from the
different perspectives of dust concentrations on its LE ellipsoid.
New LEs are reported based on difference images, and optical spectra
of these LEs are analyzed and compared.  After properly accounting for
the effects of finite dust-filament extent and inclination, we find
one field where the \ion{He}{1} $\lambda 5876$ and H$\alpha$ features
are blueshifted by an additional \about 4000~\kms\ relative to other
spectra and to the spectra of the Type IIb SN~1993J.  That same
direction does not show any shift relative to other Cas~A LE spectra
in the \ion{Ca}{2} near-infrared triplet feature.  We compare the
perspectives of the Cas~A LE dust concentrations with recent
three-dimensional modeling of the SN remnant (SNR) and note that the
location having the blueshifted \ion{He}{1} and H$\alpha$ features is
roughly in the direction of an Fe-rich
outflow and in the opposite direction of the motion of the compact
object at the center of the SNR.  We conclude that Cas~A was an
intrinsically asymmetric SN. Future LE spectroscopy of this
object, and of other historical SNe, will provide additional insight into
the connection of explosion mechanism to SN to SNR, as well as
give crucial observational evidence regarding how stars explode.
\end{abstract}

\keywords{ISM: individual (Cas~A) --- supernovae: general --- 
supernova remnants}

\section{Introduction}
\label{sec:intro}


Light echoes (LEs) are the scattered light of a transient event that
reflects off dust in the interstellar medium. The extra path length of
the two-segment trajectory results in LE light arriving at an observer
significantly later than the undelayed photons. Such circumstances
provide exciting scientific opportunities that are extremely rare in
astronomy --- specifically, to observe historical events with modern
instrumentation and to examine the same event from different lines of
sight (LoS).  We have previously employed LEs to take advantage of the
time delay, identifying systems of such LEs associated with
several-hundred-year old supernova (SN) remnants (SNRs) in the Large
Magellanic Cloud and subsequently spectroscopically classifying the
supernovae \citep[SNe;][]{Rest05b, Rest08a}. Our work represented the
first time that the spectral classification of SN light was
definitively linked with a SNR.  The analysis of the X-ray spectrum of
that SNR by \citet{Badenes08} provided confirmation of both its
classification (Type Ia) and subclass (high luminosity).  Clearly, LE
spectroscopy is a powerful technique for understanding the nature of
SNe in the Milky Way and Local Group galaxies.

Cas~A is the brightest extrasolar radio source in the sky
\citep{Ryle48} and the youngest (age \about 330~yr) Milky Way
core-collapse SNR \citep{Stephenson02}.
Its distance is approximately 3.4~kpc.  Dynamical measurements of the
SNR indicate that the explosion occurred in year $1681 \pm 19$
\citep{Fesen06_expansionasym_age}; we adopt this date for age
calculations in this paper.  A single historical account of a sighting
in 1680 by Flamsteed is attributed to the Cas~A SN \citep{Ashworth80},
although this has been disputed \citep{Kamper80}.  Cas~A is the
youngest of the certain historical CC~SNe and is thus an excellent target
for LE studies.

\citet{Krause05} identified a few moving Cas~A features (called
``infrared echoes'') using infrared (IR) images from the {\it Spitzer
  Space Telescope}, the result of dust absorbing the SN light, warming
up, and reradiating light at longer wavelengths.  Their main scientific
conclusion, that most if not all of these IR echoes were caused by a
series of recent X-ray outbursts from the compact object in the Cas~A
SNR, was incorrect because they did not take into account that the
apparent motion strongly depends on the inclination of the scattering
dust filament \citep{Dwek08, Rest11_casamotion}.  Rather, the echoes
must have been generated by an intense and short burst of ultraviolet
(UV) radiation associated with the breakout of the SN shock through
the surface of the progenitor star of Cas~A \citep{Dwek08}.

The first scattered LEs of Galactic SNe associated with Tycho's SN and
the Cas~A SN were discovered by \citet{Rest07, Rest08b}.
Contemporaneously, \citet{Krause08a} obtained a spectrum of a
scattered optical light echo spatially coincident with one of the
Cas~A IR echoes, and identified the Cas~A SN to be of Type IIb
from its similarity to the spectrum of SN~1993J,
the very well-observed and prototypical example of the SN~IIb class
\citep[e.g.,][]{Filippenko93, Richmond94, Filippenko94, Matheson00}.

\begin{figure*}[t]
\begin{center}
\epsscale{1.15}
\plotone{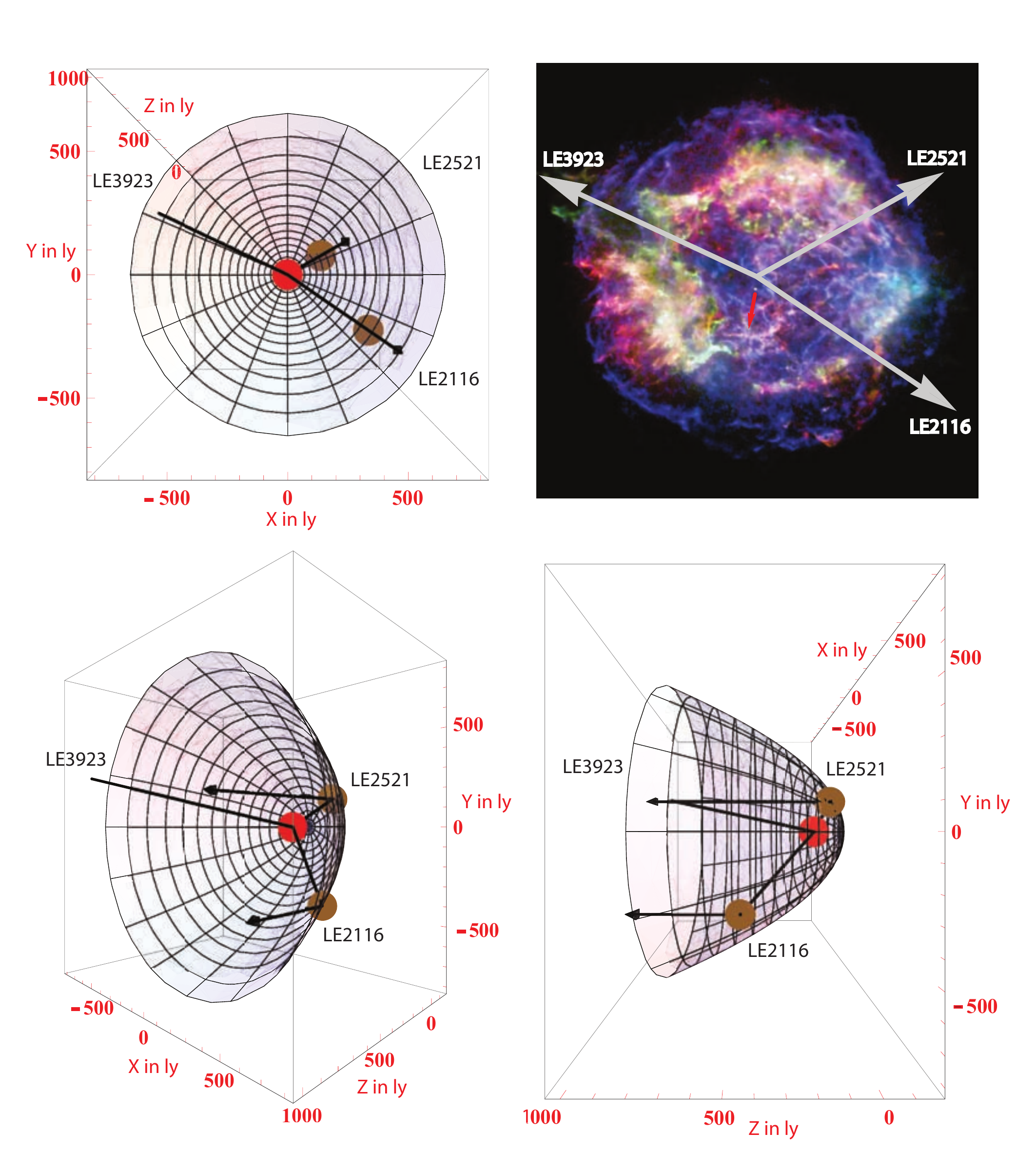}
\caption[]{North is toward the positive-$y$ axis (up), east is toward
the negative-$x$ axis (left), and the positive-$z$ axis points toward the
observer with the origin at the SNR. The SN to LE cone apex distance,
in light years, is half the interval of time since the SN
explosion. The red and brown circles indicate the SN and scattering
dust, respectively. The black lines show the path of the light
scattering from the LE-producing dust concentrations. The scattering
dust of LE3923 is more than 2000~ly in front of Cas~A, much farther
than any other scattering dust; thus, we show only part of the light
path, and do not include its dust location. The top-left panel shows an
\textit{Chandra} X-ray image \citep{Hwang04}, with the projected light
path from SN to scattering dust overplotted (gray arrows).  The red
arrow indicates the X-ray compact object and its apparent motion.
In this false-color image, red corresponds to low-energy X-rays around
the Fe~L complex (\about 1 keV and below), green to mid-energy X-rays around
the Si~K blend (\about 2 keV), and blue to high-energy X-rays in the 4--6
keV continuum band between the Ca~K and Fe~K blends.
\label{fig:3Dspecillustration}}
\end{center}
\end{figure*}

The discovery and spectroscopic follow-up observations of different
LEs from the same SN allow us to benefit from the unique advantages of
LEs --- their ability to probe the SN from significantly different
directions. The only time this technique has been applied previously
was by \citet{Smith01_eta,Smith03_eta}, who used spectra of the
reflection nebula of $\eta$~Carinae to observe its central star from
different directions.  Dust concentrations scattering SN light lie at
numerous different position angles and at different radial distances
from the observer.  This is illustrated in
Figure~\ref{fig:3Dspecillustration}, which shows the Cas~A SNR (red
dots), the scattering dust (brown dots), and the light paths.  We
denote the light echoes as LE2116, LE2521, and LE3923, where the
number represents the ID of the grid tile of the search area in this
region of the sky.  LE2521 and LE3923 are newly discovered and LE2116
was discovered by \citet{Rest08b}.  The scattering dust of LE3923 is
more than 2000~ly in front of Cas~A, much farther than any other
scattering dust; thus, we show only part of its light path.


The study of a single SN from different LoS is particularly relevant
for the Cas~A SNR.  With observations of the LEs shown in
Figure~\ref{fig:3Dspecillustration}, we are positioned to {\it
  directly} measure the symmetry of a core-collapse SN, and compare it
to the structure of the remnant.  The thermal X-ray emission as well
as the optical emission of the Cas~A SNR is very inhomogeneous, with
large Fe-rich and Si-rich outflows being spatially distinct,
indicating that the SN explosion was asymmetric
\citep{Hughes00,Hwang04,Fesen06_expansionasym_age}. The nature of
these outflows and their individual relevance to the SN explosion is
still debated \citep{Burrows05,Wheeler08,Delaney10}.
\citet{Tananbaum99} detected in \textit{Chandra} images a compact
X-ray source 7\arcsec\ from the Cas~A SNR center; it is an excellent
candidate for being the neutron star produced by the SN explosion
\citep{Tananbaum99,Fesen06_CO}. 
The position angle of the X-ray source is  off by only  $\sim
30$\arcdeg\ from the position angle of the southeast Fe-rich structure
\citep{Wheeler08}, and has a projected apparent motion of 350~\kms.
In \S \ref{sec:observations} of this paper, we report two new LE 
complexes associated with Cas~A, and we show optical spectra of 
these LEs as well as one of the previously known LEs. With these
data we are able to view three distinct directions, where each dust
concentration probes different hemispheres of the SN photosphere.  In
\citet{Rest11_leprofile} we have introduced an innovative technique
for modeling both astrophysical (dust inclination, scattering, and
reddening) and observational (seeing and slit width) effects to
measure the light-curve weighted window function which is the
determining factor for the relative time-weighting of the observed
(integrated) LE spectrum.  We apply this technique in
\S \ref{sec:spectempl} to similar well-observed SNe to produce
appropriate comparison spectra.  Such spectra are necessary to compare
the Cas~A spectra to other SNe as well as to compare the LE spectra to
each other.  We show that the Cas~A SN was indeed very similar to the
prototypical Type~IIb SN~1993J, as claimed by \citet{Krause08a}.  In
\S \ref{sec:speccomp} we demonstrate that, despite the excellent
agreement between the Cas~A spectra and SN~1993J, one LE has a
systematically higher ejecta velocity than either SN~1993J \textit{or
  the other LEs}, revealing that Cas~A was an intrinsically asymmetric
explosion; observers from different directions would have viewed a
``different'' SN spectroscopically.  In this section we also discuss
the implications of our finding for both other historical SNe and for
core-collapse SNe and their explosions.

\section{Observations \& Reductions}
\label{sec:observations}

\subsection{Imaging}

We reobserved LEs discovered in a campaign of several observing runs
on the Mayall 4~m telescope at Kitt Peak National Observatory (KPNO)
starting in 2006 \citep{Rest08b}. As described in that paper, the
Mosaic imager, which operates at the $f$/3.1 prime focus at an
effective focal ratio of $f$/2.9, was used with the Bernstein $VR$
broad-band filter (k1040) which has a central wavelength of
5945~\AA\ and a full width at half-maximum intensity (FWHM) of
2120~\AA.  The images were kernel- and flux-matched, aligned,
subtracted, and masked using the SMSN pipeline \citep{Rest05a, Garg07,
  Miknaitis07}.  LE2116 had been previously discovered and reported
\citep{Rest08b}, whereas LE2521 and LE3923 are light echoes discovered
on 2009 September 14 and 16, respectively (UT dates are used throughout 
this paper).

\begin{figure}[t]
\begin{center}
\epsscale{1.15}
\plotone{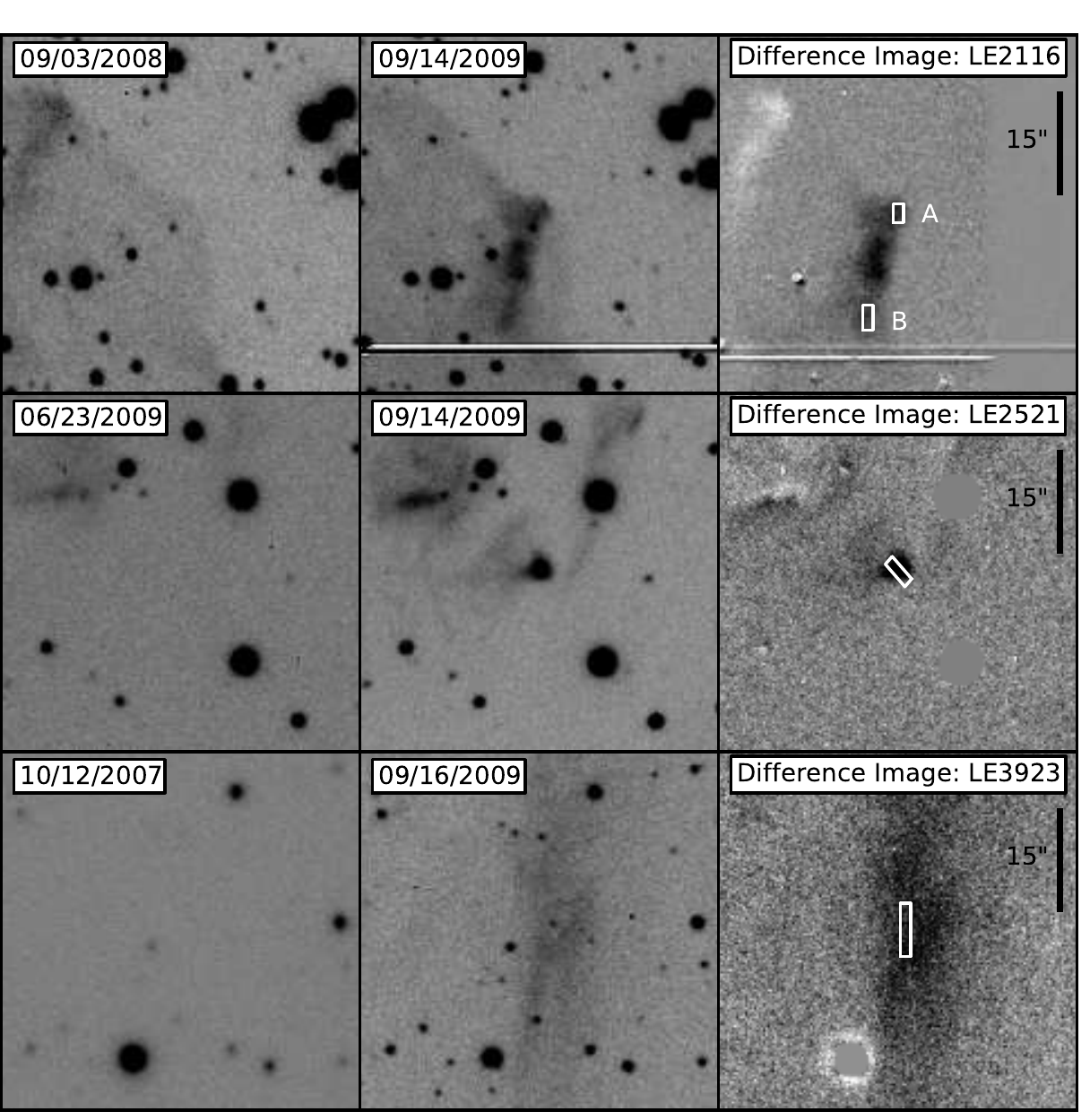}
\caption[]{Image stamps of the LE discovery images and slit positions
for spectroscopy for the three LE complexes LE2116, LE2521, and LE3923
(along horizontal panels).  Left vertical panels are the reference
template image at an earlier epoch (on the date shown in the box at top).
Middle vertical panels are the image fields at a later epoch, centered
here on the LE position.  Right vertical panels are the difference
images of the respective recent image and the earlier reference
template to isolate the transient flux of the LE. For all image stamps
north is up and east is to the left.  The position and size of the
slit used for spectroscopy is indicated by the rectangular overlay.
\label{fig:substamps}}
\end{center}
\end{figure}

Figure~\ref{fig:substamps}
shows the LEs and the adopted spectroscopic slit positions.
For the LE2116 mask design, we used images from KPNO obtained on 2009
September 14, \about 1~week before the spectroscopy was obtained.
Note that we had a third slit in the high surface brightness region
between slits A and B. Unfortunately, the LE filled the entire slit,
making proper sky subtraction impossible.  Sky subtraction using the
sky observed in other slits produced inaccurate spectra with large
residuals near the sky lines.  As a result, we have not included the
spectra from this slit in our analysis.
For LE3923, we
used images obtained at the Apache Point Observatory 3.5~m telescope
on 2009 October 16 with the
SPIcam\footnote{http://www.apo.nmsu.edu/arc35m/Instruments/SPICAM/.}
CCD imager in the SDSS-$r'$ filter and processed in a similar manner
with the SMSN pipeline.  Since the apparent motion of the LEs is
\about 30\arcsec\ per year, we measured the apparent motion and
adjusted the slit position accordingly.  These position adjustments
amounted to \about 1\arcsec.  LE2521 was sufficiently compact and
bright that long-slit spectroscopy was favored over masks.  
A summary of the spectroscopic and geometric parameters is given in
Tables \ref{tab:specinfo}~and~\ref{tab:geoinfo}, respectively.  For
the light-echo profile fits described in \S \ref{sec:spectempl}, we
used the deep images in good seeing from 2009 September 14--16.
\begin{deluxetable*}{cccccccccccc}
\tabletypesize{\scriptsize}
\tablecaption{
\label{tab:specinfo}}
\tablehead{
\colhead{} &
\colhead{} &
\colhead{R.A.} &
\colhead{decl.} &
\colhead{PA SNR-LE} &
\colhead{} &
\colhead{Seeing\tablenotemark{a}} &
\colhead{} &
\colhead{Width\tablenotemark{a}} &
\colhead{Length\tablenotemark{a}} &
\colhead{PA\tablenotemark{a}} &
\colhead{} \\
\colhead{LE} &
\colhead{Telescope} &
\colhead{(J2000)} &
\colhead{(J2000)} &
\colhead{(\arcdeg)} &
\colhead{UT Date\tablenotemark{a}} &
\colhead{(\arcsec)} &
\colhead{ID\tablenotemark{a}} &
\colhead{(\arcsec)} &
\colhead{(\arcsec)} &
\colhead{(\arcdeg)} &
\colhead{UT Mask\tablenotemark{b}}
}
\startdata
 2116 & Keck & 23:02:27.10 & +56:54:23.4 & 237.79           & 20090922 & 0.81 & A       & 1.5 & 2.69 & \phantom{0}0.0 & 20090914 \\
 2116 & Keck & 23:02:27.67 & +56:54:07.7 & 237.79           & 20090922 & 0.81 & B       & 1.5 & 3.59 & \phantom{0}0.0 & 20090914 \\
 2521 & Keck & 23:12:03.86 & +59:34:59.3 & 299.10           & 20090922 & 0.79 & \nodata & 1.5 & 4.32 &           41.0 & \nodata  \\
 2521 & MMT  & 23:12:03.88 & +59:34:59.2 & 299.10           & 20090921 & 0.70 & \nodata & 1.0 & 4.20 &           30.0 & \nodata  \\
 3923 & Keck & 00:21:04.97 & +61:15:37.3 & \phantom{0}65.08 & 20091023 & 0.89 & \nodata & 1.5 & 7.72 & \phantom{0}0.0 & 20091016
\enddata
\tablenotetext{a}{Parameters for spectroscopic slit.}
\tablenotetext{b}{UT date of image used for mask design.}
\end{deluxetable*}
%
%
\begin{deluxetable}{ccccccccc}
\tabletypesize{\scriptsize}
\tablecaption{
\label{tab:geoinfo}}
\tablehead{
\colhead{} &  \colhead{$\delta$\tablenotemark{a}} & \colhead{$\rho$\tablenotemark{b}} & \colhead{$x$\tablenotemark{c}} & \colhead{$y$\tablenotemark{c}} & \colhead{$z$\tablenotemark{c}}\\
\colhead{LE} &  \colhead{(\arcdeg)} & \colhead{(ly)} & \colhead{(ly)} & \colhead{(ly)} & \colhead{(ly)}
}
\startdata
  2116 &      3.376 &  \phantom{1}628 &   \phantom{-1}531 &  -335            &   \phantom{2}437\\
  2521 &      1.642 &  \phantom{1}318 &   \phantom{-1}278 &   \phantom{-}155 &   \phantom{2}-10\\
  3923 &      7.587 & 1204            &             -1092 &   \phantom{-}507 &  2045\\
\enddata
\tablenotetext{a}{Angular distance between SNR and LE.}
\tablenotetext{b}{Distance between SNR and LE in the plane of the sky ($\rho^2=x^2+y^2$).}
\tablenotetext{c}{Coordinates $x$,$y$, and $z$ of the LE with
origin at the SNR. The positive $x$ and $y$ axis are in the plane of
the sky toward West and North, respectively. The positive $z$ axis is
along the line of sight from the SNR toward the observer. We assume a
distance to the Cas A SNR of 3.4~kpc and 1681 AD as the time of
explosion.}
\end{deluxetable}

\subsection{Spectroscopy}

We obtained spectra of the LEs with the Low Resolution Imaging
Spectrometer \citep[LRIS;][]{Oke95} on the 10~m Keck~I telescope and
with the Blue Channel spectrograph \citep{Schmidt89} on the 6.5~m MMT
(see Table~\ref{tab:specinfo}).  For the LRIS observations, slit masks
were designed to maximize the efficiency of the telescope time.
Standard CCD processing and spectrum extraction were performed with
IRAF\footnote{IRAF: the Image Reduction and Analysis Facility is
distributed by the National Optical Astronomy Observatory, which is
operated by the Association of Universities for Research in Astronomy
(AURA), Inc., under cooperative agreement with the National Science
Foundation (NSF).}.  The data were extracted using the optimal
algorithm of \citet{Horne86}.  Low-order polynomial fits to
calibration-lamp spectra were used to establish the wavelength scale,
and small adjustments derived from night-sky lines in the object
frames were applied.  We employed our own IDL routines to flux
calibrate the data and remove telluric lines using the well-exposed
continua of spectrophotometric standard stars \citep{Wade88,
Matheson00, Foley03}.  For LE2116, we combine slits A and B using
weights of 0.9 and 0.1, respectively.  We present our LE spectra in \S
\ref{sec:speccomp}.


\section{Generating Comparison Spectra}
\label{sec:spectempl}

\begin{figure*}[t]
\epsscale{1.15}
\plottwo{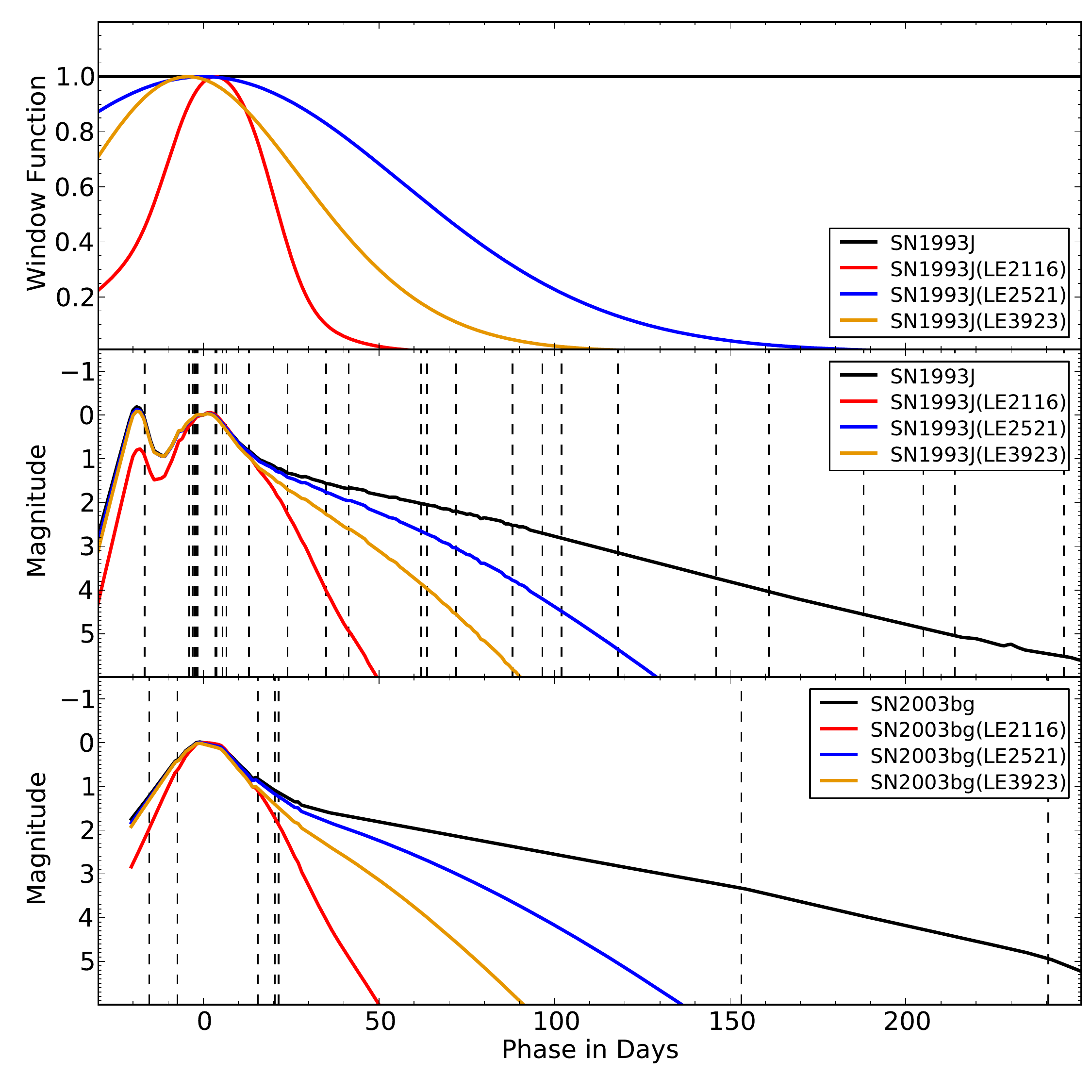}{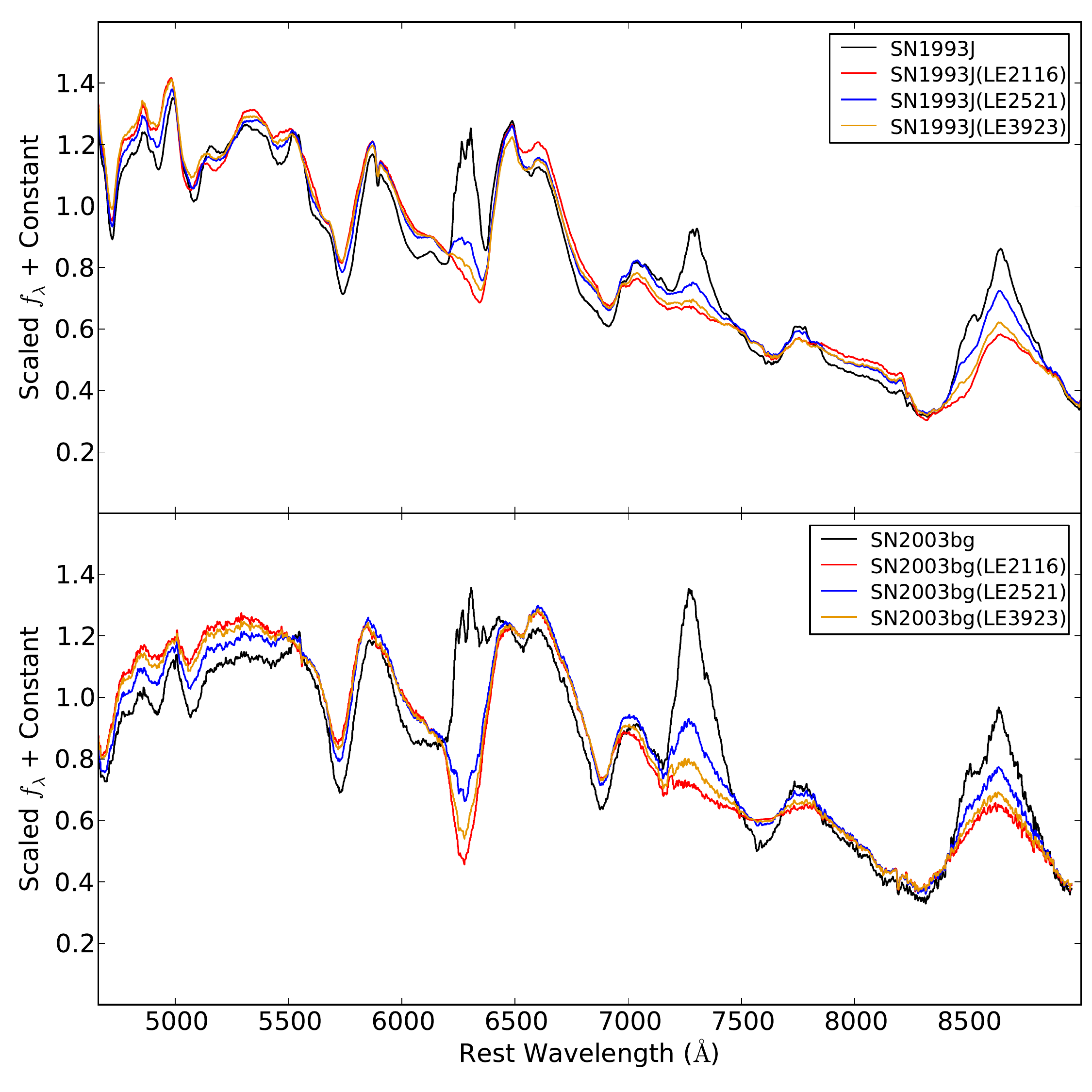}
\caption[]{Red, cyan, and blue indicate LEs LE2521, LE3923, and
  LE2116, respectively. \emph{Top left:} window function for each
  LE. \emph{Middle left:} effective light curves of SN~1993J for each
  LE. The unmodified light curve of SN~1993J is shown in black. Epochs
  with spectra are indicated by the dashed lines. \emph{Bottom left:}
  same as middle left, but for SN~2003bg. \emph{Top right:} integrated
  SN~1993J spectra, where the integration is weighted by the
  respective effective light curve. Note that the black spectrum is
  weighted by the original, unmodified SN~1993J light curve.  The
  differences in the spectra are mainly due to the strong lines of
  [\ion{O}{1}] $\lambda\lambda 6300$, 6363, [\ion{Ca}{2}]
  $\lambda\lambda 7291$, 7324, and the \ion{Ca}{2} NIR
  triplet in the late-phase spectra.  \emph{Bottom right:} same as top
  right, but for SN~2003bg.
\label{fig:w_efflc}}
\end{figure*}

We have recently determined that modeling the dust-filament properties
(e.g., the dust width) is required to accurately model LE spectra
\citep{Rest11_leprofile}.  All previous studies have neglected this
aspect. Because the dust
filament that reflects the LE has a nonzero size, it translates into a
window function over the time domain as the SN light traverses over
it.  If this window function is narrower than the SN light curve, the
observed LE spectrum is affected.  We find that the observed LE
spectrum is a function of dust-filament thickness, dust-filament
inclination, seeing, spectrograph slit width, and slit
rotation/position with respect to the LE. All of the noninstrumental
parameters can be determined with an analysis of the LE in images at
different epochs, and an individually modeled spectral template for a
given observed LE can be constructed {\it a priori} without using the
actual observed LE spectrum. \citet{Rest11_leprofile} describe this 
process in detail, showing that the observed LE spectrum is
the integration of the individual spectra weighted with an effective
light curve, which is the product of the light curve with the
dust-filament-dependent window function. We emphasize that these
window functions are different for every LE location. The top-left
panel of Figure~\ref{fig:w_efflc} shows the window functions for the
LEs for which we have spectra using the SN~1993J spectral library
\citep{Jeffery94,Barbon95,Fransson05} and light curve
\citep{Richmond96}. We refer the reader to \citet{Rest11_leprofile}
for a detailed description of how the effective light curves are
derived.  Examples of the fitted parameters are given in
Tables~1~and~2 and shown in Figure~12 of \citet{Rest11_leprofile}

Even though LE2521 and LE2116 have comparable dust widths, LE2521 has
a considerably wider window function than LE2116 since its
dust-filament inclination ($\alpha=54\arcdeg$) is significantly larger
than that of LE2116 ($\alpha=9\arcdeg$), causing the projected SN
light-curve shape to be ``squashed.''  LE3923 has an inclination
similar to that of LE2116, but its scattering dust is thicker, leading
to a wider window function.  We note that the dust-filament
inclination and width are the two most influential parameters
affecting the window function.


The middle-left and bottom-left panels of Figure~\ref{fig:w_efflc} show 
the effective light curves of SN~1993J and SN~2003bg, respectively. 
We constructed the effective light curve of SN~2003bg using the
spectra and light curve in \citet{Hamuy09}.
In general, the rise time of a SN is shorter than 50 days. Since the
window functions have, in most cases, FWHM $\ge 30$~days (see the
top-left panel in Figure~\ref{fig:w_efflc}), the rising wing of the
effective light curves and the real light curve are similar (see the
middle-left and bottom-left panels in Figure~\ref{fig:w_efflc}), and
consequently the impact on the integrated spectra is small. However,
the decline time of a SN is significantly longer, and therefore the
window function has a much more profound impact at late phases: it
cuts off the contribution of spectra with phases $\ge 100$~days.  Even
though the intrinsic brightness of the SN at late phases is much
fainter than at peak, these late-phase spectra still significantly
contribute to the integrated spectra since they are completely
dominated by a few persistent lines. This is illustrated in the top-right
and bottom-right panels of Figure~\ref{fig:w_efflc}, which show the
integrated spectra of SN~1993J and SN~2003bg, respectively.  The
late-time spectra of SNe~IIb such as SN~1993J and SN~2003bg are
dominated by [\ion{O}{1}] $\lambda\lambda 6300$, 6363, [\ion{Ca}{2}]
$\lambda\lambda 7291$, 7324, and the \ion{Ca}{2} NIR triplet. Note
that the black spectra, which are weighted by the original,
unmodified light curve, are much stronger at these lines compared to
the other spectra, which are integrated using the effective
light curves. This figure clearly illustrates the impact and importance
of the effective light curve on the contribution of the late-phase
spectra to the observed integrated spectrum. 

\subsection{Integrating and Fitting the Spectral Templates to the 
Observed LE Spectra}

We numerically integrate the spectral templates of SN~1993J
\citep{Jeffery94,Barbon95,Richmond96,Fransson05}, SN~2003bg
\citep{Hamuy09}, and SN~2008ax \citep{Chornock_08ax} by weighting the
individual epochs with the light curve as described by
\citet{Rest08a}, with the only difference being that the real light
curve is replaced with an effective light curve as described in \S
\ref{sec:spectempl} and in more detail by \citet{Rest11_leprofile}.

\begin{figure}
\begin{center}
\epsscale{1.15}
\plotone{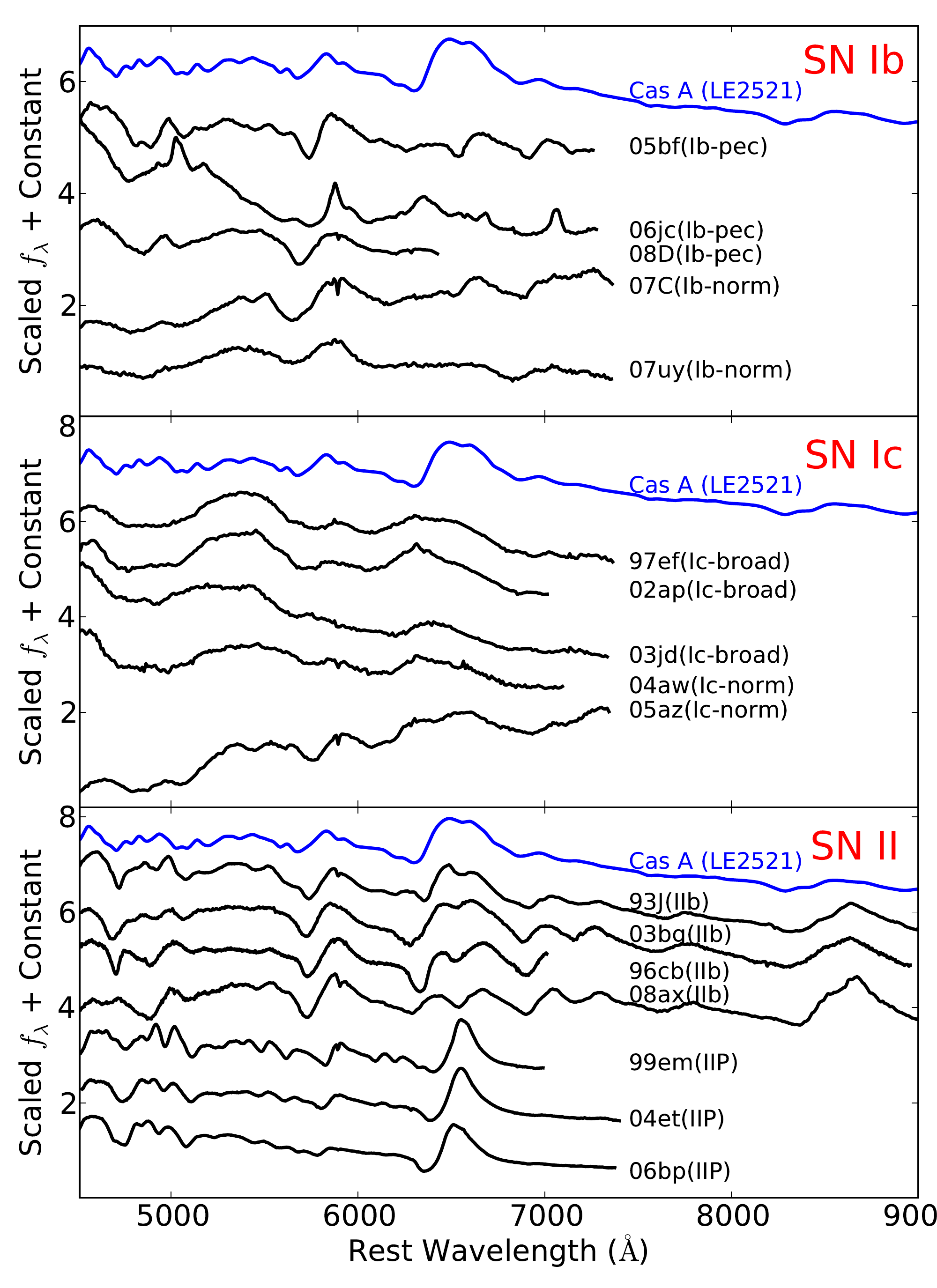}
\end{center}
\caption[]{The LE spectrum of LE2521 (blue line)
compared to various SNe~Ib, Ic, and II using the window function shown in 
Figure~\ref{fig:w_efflc}. SN~1993J and SN~2003bg unambiguously give the 
best fit.
\label{fig:comp2allSN}}
\end{figure}

Several authors have addressed the single-scattering approximation for
LEs \citep[e.g.,][]{Couderc39, Chevalier86, Emmering89, Sugerman03,
  Patat05}. The total surface brightness depends on parameters such as
the intrinsic brightness of the event, dust density and thickness, and
others.  For our purposes, we only need to consider the parameters
that attenuate the spectrum, which are (1) the forward scattering
described by the integrated scattering function $S(\lambda,\theta)$ as
derived by \citet{Sugerman03}, and (2) the reddening due to the
extinction by Galactic dust. The attenuation by forward scattering can
be determined {\it a priori} using the observed angular separation of
the LE from Cas~A.  The only uncertainties introduced here are due to
the assumed age and distance for Cas~A. The reddening cannot be
determined independently and therefore we fit for it. We fit the
template spectra to the observed LE spectra with the only free
parameters being the normalization and the
reddening. Figure~\ref{fig:comp2allSN} compares the LE spectrum of
LE2521 (blue line) to various SNe~Ib, Ic, and II using
the window function for LE2521.  SN~1993J and SN~2003bg
quite clearly provide the best fit.


\section{Comparison of the Cas~A Light Echoes}
\label{sec:speccomp}

\begin{figure*}
\begin{center}
\epsscale{1.15}
{
\plotone{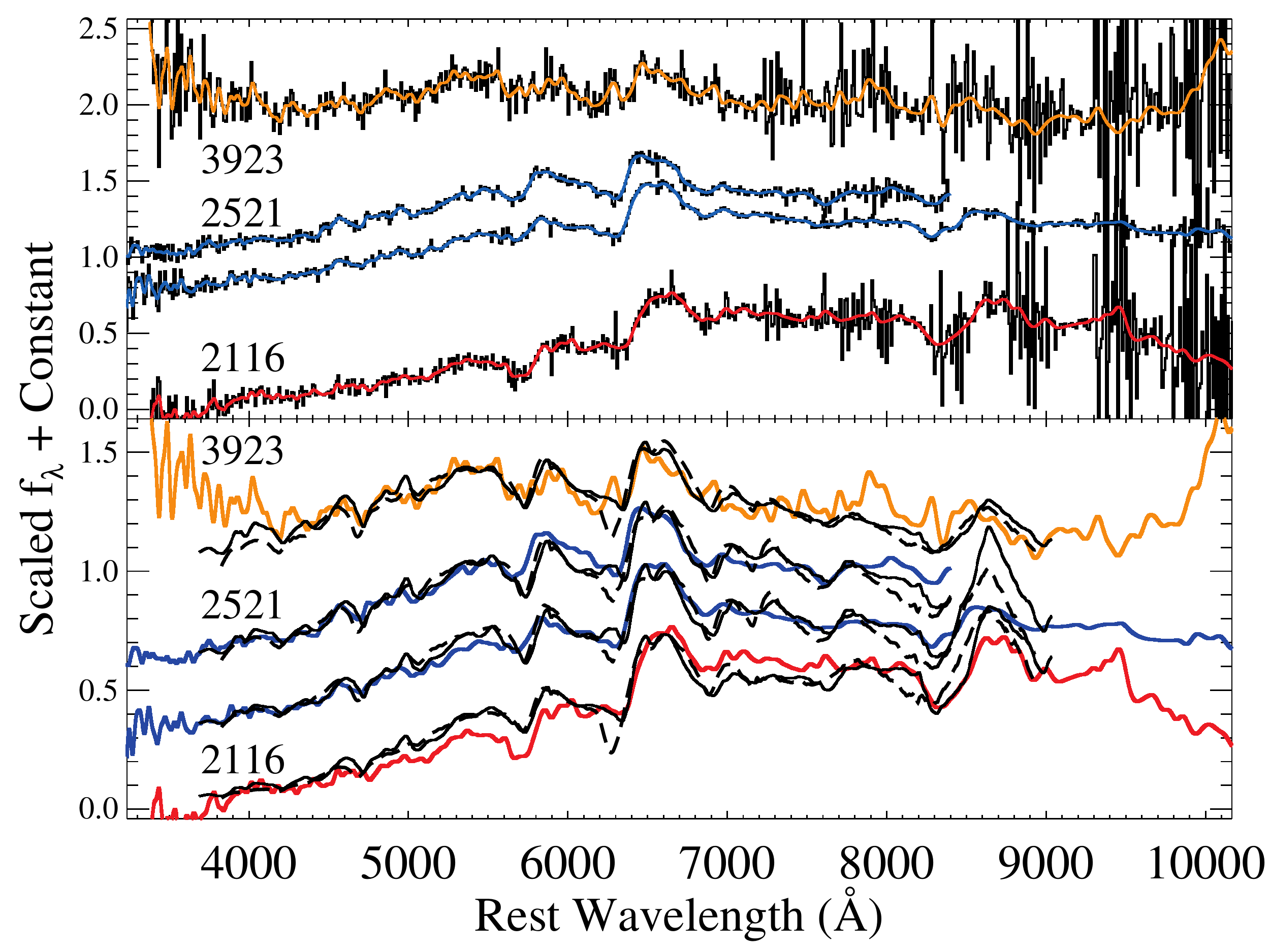}}
\end{center}
\caption[]{\emph{Top panel}: observed LE spectra binned to
  10~\AA\ pixel$^{-1}$ (black curves).  The colored curves are the
  variance-weighted smoothed spectra with the red, blue, and orange
  curves corresponding to LEs LE2116, LE2521, and LE3923,
  respectively.  The LE2521 spectrum with the shorter wavelength range
  is from the MMT; the remaining spectra are from Keck.  \emph{Bottom
  panel}: variance-weighted smoothed LE spectra (colored curves)
  compared to their corresponding light-curve weighted integrated
  SN~1993J (solid black curves) and SN~2003bg spectra (dashed black
  curves).  The LE spectra are equivalent to the spectra in the top
  panel.  The SNe~1993J and 2003bg spectra have been processed to
  reflect the dust scattering and reddening associated with each LE
  (see text for details).  The spectra of Cas~A are very similar to
  both SNe~1993J and 2003bg from all three directions probed; however,
  there are some significant differences (see \S \ref{sec:speccomp}).
\label{fig:observedspectra}}
\end{figure*}

In Figure~\ref{fig:observedspectra}, we present four spectra of Cas~A
LEs, corresponding to three different directions.  LE2521 was much
brighter than the other LEs, and as a result, its spectrum has a
higher signal-to-noise ratio (S/N) than the others.  The spectrum of
LE3923 has the lowest S/N, but it still displays the same spectral
features as the other spectra.  All three directions have very similar
spectra with the main differences being the continuum shape, which we
have shown to be a result of different scattering and reddening for
the various directions.  The line profiles do exhibit some
differences, but as shown in \S \ref{sec:spectempl}, different window
functions can create significant differences in the spectra.

To properly compare the LEs from different directions, one must make a
comparison to another SN, which removes the effect of the window
function.  Assuming that the window functions are correct, if a LE
spectrum is the same as the integrated spectrum of a SN, then the two
objects likely looked the same at maximum light (there is obviously
some degeneracy between the different epochs within the window
function).  If the two spectra are different, then either their
maximum-light spectra or light curves were likely different;
regardless, it is a direct indication that the two objects appeared
different along those LoS.  Moreover, if one LE matches the integrated
spectrum of a particular spectrum, but a LE from a different direction
does not, then observers along the two LoS would have seen
``different'' SNe.

For Cas~A, we have compared the LEs to spectra of SNe~1993J, 2003bg, and 
2008ax (see Fig.~\ref{fig:observedspectra}).  These are the best-observed
SNe~IIb for which we had light curves and a good time series of
spectra.  As described above, SN~1993J is an iconic SN~IIb that was
extensively observed \citep[e.g.,][]{Filippenko93, Richmond94,
  Filippenko94, Matheson00} and whose progenitor system was identified
in pre-explosion images \citep{Podsiadlowski93, Aldering94}, and its
binary companion was identified through late-time spectroscopy
\citep{Maund04}. SN~2003bg was a luminous event (both optically and in
the radio) with broad lines at early times \citep{Soderberg06,
  Hamuy09, Mazzali09}.  SN~2008ax was discovered hours after
explosion, which enabled significant follow-up observations
\citep{Pastorello08, Roming09, Chornock10}, including extensive
spectropolarimetry \citep{Chornock10} indicating that the outer layers
of the ejecta were rather aspherical.  Furthermore, its progenitor
system was also detected in pre-explosion images \citep{Crockett08}.

Examining Figure~\ref{fig:observedspectra}, it is clear that the LE
spectra in all directions are similar to each other. In addition, the
LE spectra are similar to spectra of SNe~1993J and 2003bg, indicating
that no dramatic differences are seen from the various directions
(such as a different spectral classification).  SN~2008ax is a worse
match to the LEs, having significantly weaker H$\alpha$ emission than
the other objects.  Since our spectral series of SN~2008ax (both CfA
and published spectra from \citealt{Chornock10}) is not as exhaustive
as that of SNe~1993J and 2003bg, we do not know if the differences are
intrinsic or the result of incomplete data.  Because of the poor match
and the differences not necessarily being physical, as well as for
clarity, we do not show in Figure~\ref{fig:observedspectra} the
comparison spectra generated from SN~2008ax.

\begin{figure*}
\begin{center}
\epsscale{1.15}
{
\plotone{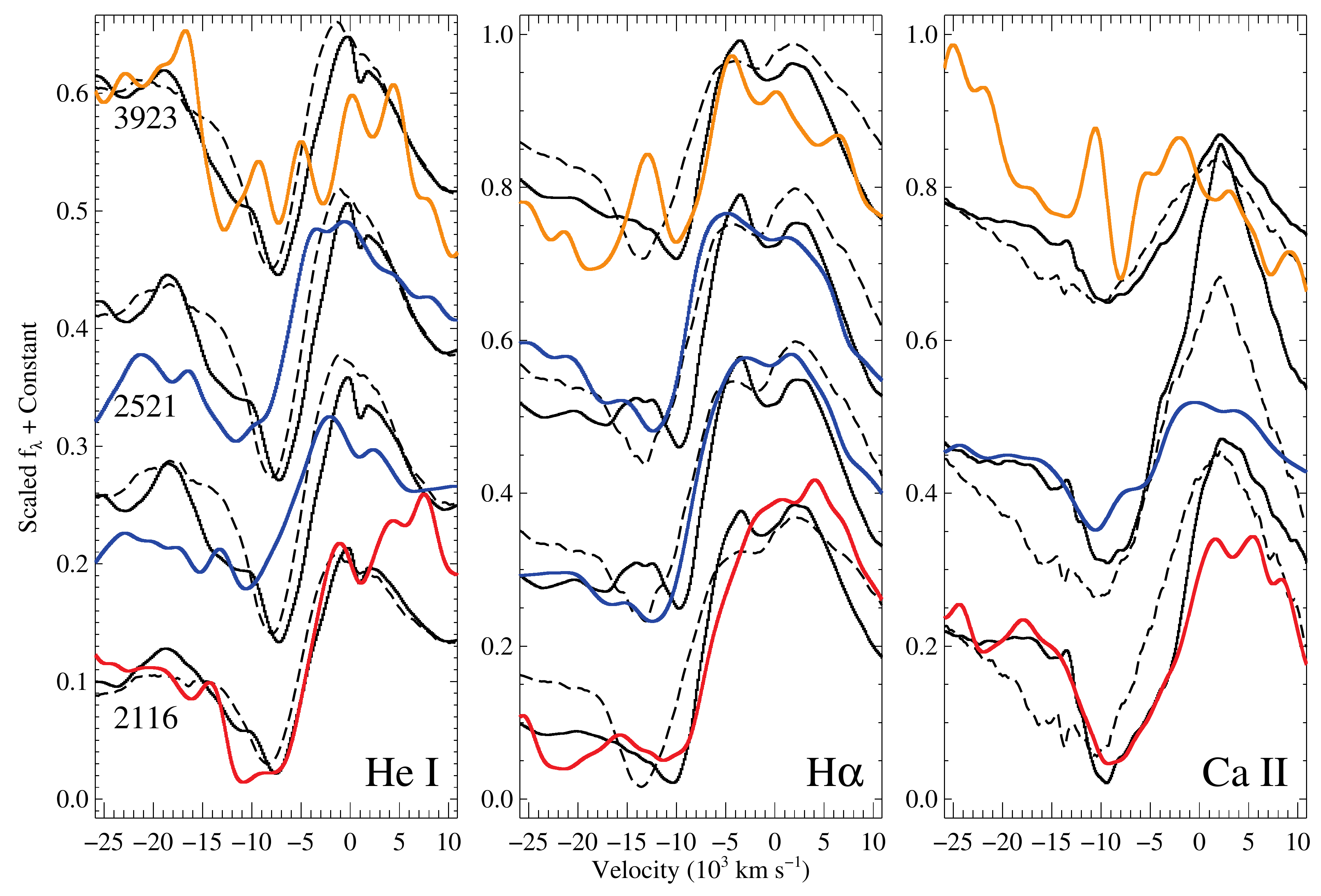}}
\caption[]{Observed LE spectra near the \ion{He}{1} $\lambda
5876$ (left), H$\alpha$ (middle), and \ion{Ca}{2} NIR triplet (right)
features, shown on a velocity scale.  The red, blue, and orange curves
correspond to LEs LE2116, LE2521, and LE3923, respectively
(see Fig.~\ref{fig:observedspectra} for details).  The top LE2521
spectrum (which is not shown in the right-hand panel) is from the MMT; the
remaining spectra are from Keck.  The solid and dashed black curves
are the corresponding light-curve weighted integrated spectra of 
SNe~1993J and 2003bg, respectively.
\label{fig:casa_vel}}
\end{center}
\end{figure*}

Despite all LEs being quite similar, a detailed comparison of the LE
line profiles to those of SN~1993J in Figure~\ref{fig:casa_vel}
indicates that while two directions (LE2116 and LE3923) are consistent
with the line velocities of SN~1993J, one direction (LE2521) has
higher-velocity absorption for the \ion{He}{1} $\lambda 5876$ and
H$\alpha$ features.  Specifically, the minima of the features are
blueshifted by an additional 4000 and 3000~\kms, respectively.  In
addition to the blueshifted absorption minima, the entire profile of
the features seems to be shifted blueward, including the emission
component of the P-Cygni profile. The \ion{Ca}{2} NIR triplet may 
have a slightly larger velocity for LE2521 than in the other 
directions, but this does not appear to be significant. We estimate 
that the uncertainties in these lines are small and not 
significantly more than \about 100~\kms.

Interestingly, the comparison spectra from SN~2003bg show H$\alpha$
velocities that are larger than those of Cas~A for LE2116 and LE3923,
but the velocity matches that of Cas~A for LE2521.  For this
direction, the profile shape is also matched.  The LE2521 \ion{He}{1}
velocity (and full profile) appears to be larger than that of
SN~2003bg, which has a similar velocity to that of SN~1993J.  This
indicates that although SN~2003bg is an excellent match to the
H$\alpha$ feature of LE2521, there are still differences, and
SN~2003bg is not a perfect analog.  SN~2003bg has a slightly higher
velocity for the \ion{Ca}{2} NIR triplet than SN~1993J, which makes it
more consistent with LE2521, but we reiterate that the difference is
not significant.

%

%

\section{Comparison of the Light Echoes to the SNR structure}
\label{sec:SNR3D}


The structure of the Cas A SNR is very complex --- from the large scales,
which show spatially distinct Fe-rich and Si-rich regions, to the
small scales, which are dominated by many filaments and bright
knot-like features in the optical and X-ray bands. Overall, the ejecta
emission from the SNR appears very turbulent and inhomogeneous,
similar to other remnants of core-collapse SNe \citep[e.g.,
  G292.0+1.8,][]{Park07}. \citet{Hughes00} argue that the ejecta have
overturned during the explosion, since the Fe-rich material is
ahead of the Si-rich material in the southeast \citep[SE; however,
  see][]{Delaney10}.  The Si-rich regions in the northeast (NE) and
southwest (SW) show structure that was interpreted by \citet{Hwang04}
as indicative of a bipolar jet system. These ``jet-like'' outflows
were also seen by \citet{Fesen06_expansionasym_age} in the {\it Hubble
  Space Telescope (HST)} census of fast-moving optical knots of the
intermediate-mass elements \ion{N}{2} and \ion{O}{2}, further
supporting this NE--SW bipolar jet interpretation.

While this ``jet-counterjet'' structure is clearly a directional
outflow associated with the ejecta \citep[and not the result of some
  pre-existing bipolar cavity in the circumstellar medium;
  e.g.,][]{Blondin96}, it seems that the outflow was not energetic
enough to power the entire explosion \citep{Laming06}.  In addition,
the projected motion of the compact object detected in the X-rays
\citep{Tananbaum99, Fesen06_CO}, thought to be the neutron star or
black hole created by the explosion, is perpendicular to the NE--SW
outflows. Most jet-powered explosion models predict that the
neutron-star kick would be roughly aligned with the jet axis.
Furthermore, the Fe-rich outflows are concentrated into two regions in
the SE and NW \citep{Hughes00, Willingale02, Dewey07}, with the SE
outflow blueshifted and the NW outflow redshifted
\citep{Willingale02}.  This has given rise to a new picture that there
is an axis in the Cas~A SNR from the SE to the NW at a position angle
of \about 125\arcdeg\ aligned with the Fe-rich knots.  In this model,
the Si-rich structures to the NE and SW are only secondary features
caused by instability-powered flows from an equatorial torus
\citep{Burrows05, Wheeler08}.

In a recent study, \citet{Delaney10} used the technique of Doppler
imaging to give the most complete and updated three-dimensional (3D)
model.  This model is derived from IR (\textit{Spitzer}), optical
(ground-based and \textit{HST}), and X-ray (\textit{Chandra}) data.
\citet{Delaney10} used velocity information from specific spectral
features to deproject the structure of the SNR along the radial
direction perpendicular to the plane of the sky (see their \S 2, \S 3,
and \S 4 for details of the data reduction and deprojection techniques).
Their work reveals a very complex 3D structure that can be
characterized by a spherical component, a tilted ``thick disk,'' and
multiple ejecta outflows.  In their model, the thick disk is tilted
from the plane of the sky at an angle of \about $25^\circ$ from the
E--W axis and \about $30^\circ$ from the N--S axis. This thick disk
contains all the ejecta structures, including the most prominent
outflows that show up best in X-ray Fe~K emission, but also appear in
other datasets.  Infrared emission in the [\ion{Ar}{2}] and
[\ion{Ne}{2}] lines often reveals ring-like structures, which are
sometimes seen as broken rings, at the base of these outflows.
\citet{Delaney10} note that these rings appear at the intersection
between the thick-disk structures and a roughly spherical reverse
shock.  Some of the outflows are bipolar, with oppositely directed
flows about the expansion center, while others are not. In particular,
the blueshifted emission from the Fe-rich outflow in the SE is clearly
collimated, but the corresponding redshifted emission in the NW is
not.

The 3D analysis by \citet{Delaney10} suggests that the Cas~A SN
explosion was highly asymmetric, with most of the ejecta flattened in
a thick disk that is slightly tilted from the plane of the sky and no
prominent structures perpendicular to the plane of the sky.  It is
possible that this might be due to selection effects, at least to some
extent, because limb brightening will make structures close to the
plane of the sky easier to detect, but given the quality of the
individual datasets, it is unlikely that any prominent outflows along
the radial direction would have been missed completely by
\citet{Delaney10}.

\begin{figure*}
\begin{center}
\epsscale{0.35}
\rotatebox{90}
{
\plotone{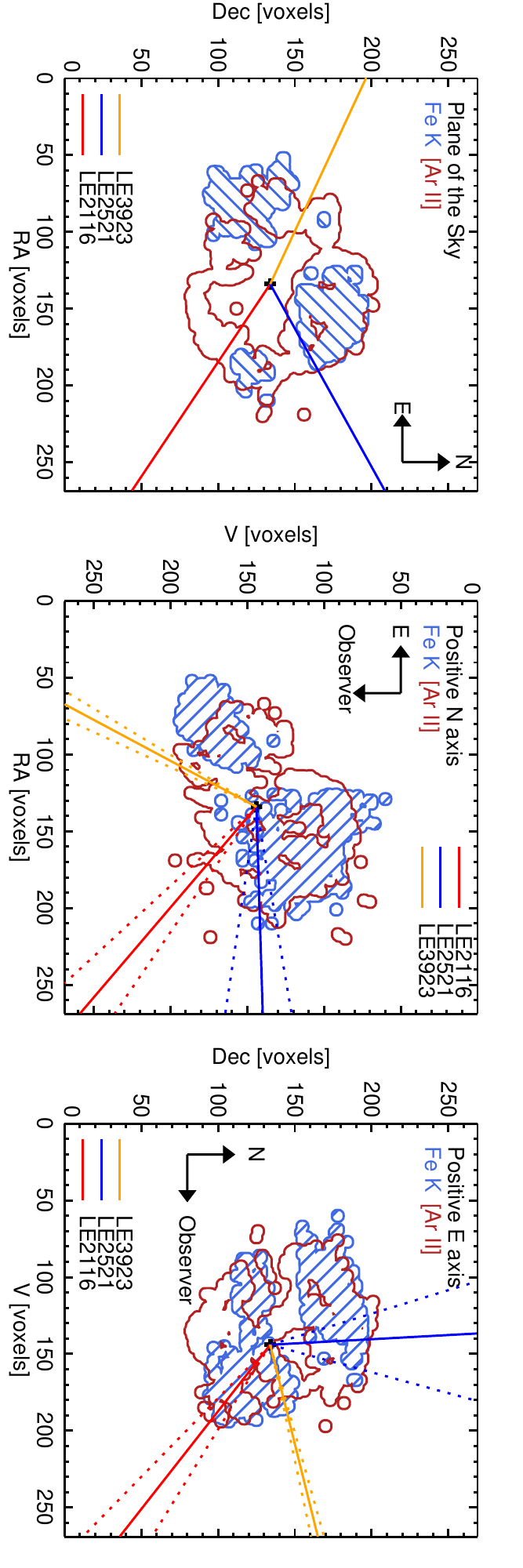}}
\caption[]{ Three-dimensional [\ion{Ar}{2}] (red) and Fe~K (blue) 
datasets from \citet{Delaney10}. The datasets are viewed from Earth's
perspective (as they appear on the plane of the sky, left panel), from
above (from a point along the positive N axis, middle panel), and
sideways (from a point along the positive E axis, right panel). The
datasets are represented in discrete voxels (volumetric pixels),
roughly $0.2''$ on a side \citep[see][for details]{Delaney10}. The
expansion center is represented by a black cross in all three
panels. The LE directions are shown for the fiducial distance of 3.4
kpc (solid lines), and for reasonable upper and lower limits,
respectively 4.0 and 3.0 kpc (dotted lines).
\label{fig:casa_3D_LEs}}
\end{center}
\end{figure*}

In Figure~\ref{fig:casa_3D_LEs}, we superimpose the LoS from our LEs
onto three of the [\ion{Ar}{2}] and Fe~K datasets from
\citet{Delaney10}.  The thick disk is apparent in the structure of the
SNR as seen from the positive N axis (middle panel of
Fig.~\ref{fig:casa_3D_LEs}), and (to a lesser extent) the positive E
axis (right panel of Fig.~\ref{fig:casa_3D_LEs}).  Among our
observed LEs, LE2116
and LE3923 are sampling LoS away from
this disk, and do not intersect any prominent ejecta structures.  The
LoS of LE2521, on the other hand, intersects the edge of the large
complex of Fe~K emission in the NW that is at the edge of the thick
disk.  

\section{Discussion}
\label{sec:discussion}

\subsection{Systematic Effects}

To properly assess differences in the LE spectra, we have
compared each LE to light-curve and window-function weighted
spectra of well-studied SNe.  When assessing the similarities of a
LE to another SN, this process is straightforward.  However,
when comparing different LEs by examining their similarities
to other SNe, there are potential systematic effects.

\subsubsection{Window Functions and Comparison Objects}

When making integrated comparison spectra, the light curve and spectra
of that particular SN are used. If that object is not a perfect analog
of Cas~A, then different window functions can introduce apparent
differences between the LEs.  As an extreme example, consider two
window functions: one which is a top hat only covering the premaximum
portion of the light curve, and another top hat only covering the
postmaximum portion of the light curve.  If the comparison SN had
lower line velocities than Cas~A before maximum brightness, but similar 
line velocities after maximum, the comparison spectra would indicate a
difference for the two LEs.

For the LEs presented here, the window functions are not significantly
different.  In Figure~\ref{fig:w_efflc}, we see that the major
differences are the contribution of the light curve 50--140~days after
maximum brightness. One main source of uncertainty in the window
function is caused by the uncertainty in the dust inclination.  The
dust inclination is derived from the apparent motion. In particular,
for faint LEs, the uncertainty in the apparent motion can
introduce uncertainties in the inclination of $\sim 5$\arcdeg\,
\citep{Rest11_leprofile}. We have not yet fully explored all ways to
improve how we determine the dust inclination, and we hope that better
modeling will further reduce its uncertainty in the future. Another
source of uncertainty is caused by imperfections in the light-echo
profile modeling and thus the dust width caused by substructure in the
dust.  
This can be tested with the SN~1987A LEs
\citep{Kunkel87,Crotts88,Suntzeff88}, since the spectrophotometric
evolution of the SN itself has been monitored extensively.  We show in
\citet{Rest11_leprofile} and \citet{Sinnott_87Aspec} that the observed
LE profile, apparent motion, and spectra can all be brought in
excellent accordance with what is predicted using the SN~1987A
spectrophotometric library.


Figure~\ref{fig:w_efflc} shows how the window functions change the
integrated spectra, including a flat window function, corresponding to
the full light-curve weighted integrated spectrum.  The most
significant differences occur at the positions of nebular emission
lines, particularly [\ion{O}{1}] $\lambda\lambda 6300$, 6363,
[\ion{Ca}{2}] $\lambda\lambda 7291$, 7324, and the \ion{Ca}{2} NIR
triplet.

The H$\alpha$ line is most affected by the [\ion{O}{1}] feature.  For
SN~1993J, as [\ion{O}{1}] becomes more prominent, it can make the
velocity of the minimum of the H$\alpha$ absorption appear lower,
which may affect our interpretation of this feature for Cas~A.
However, it does not affect the blue edge of the emission feature.
For SN~2003bg, the H$\alpha$ absorption is at a higher velocity,
coinciding almost exactly with the peak of the [\ion{O}{1}] emission.
As a result, the differences in window functions do not dramatically
change the apparent velocity of the H$\alpha$ absorption, but rather
just affect its strength.

For both SNe~1993J and 2003bg, the \ion{He}{1} feature does not change
much with different window functions.  There is a noticeable change in
the spectra between those with the measured window functions and the
full light-curve weighted spectrum.  The full light-curve weighted
spectrum has a slightly lower velocity for both SNe~1993J and 2003bg.

Although we cannot rule out that the window function has created the
differences seen in Cas~A, that scenario requires particular tuning.
One way this could happen is if Cas~A changes from a SN~1993J-like
spectrum to a higher-velocity spectrum from about 80~days after
maximum, when the window functions for LE2116 and LE3923 are 6~mag
below peak, significantly less than the window function of LE2521. In
addition, most of the flux at these late phases needs to be in
H$\alpha$ and \ion{He}{1} instead of the nebular emission lines, which
is physically unlikely for SNe~IIb. Then the integrated spectrum of
SN~1993J could show a lower velocity than LE2521 while still having
the same velocity for the other directions.  Additional comparison
objects and LE spectra with different window functions should address
this potential systematic effect.

\subsection{Connection to Explosion Models}

The LE data alone suggest that Cas~A was an asymmetric SN.  As
outlined in \S \ref{sec:SNR3D}, this is consistent with observations
of the SNR, which indicate the existence of a rich set of multiple
outflows \citep[e.g.,][]{Hughes00, Willingale02,
  Fesen06_expansionasym_age}.  The most prominent are bipolar Fe-rich
outflows in the SE and NW that are blueshifted and redshifted,
respectively.  These outflows are approximately (but not exactly)
aligned along a single axis through the kinematic center of the SNR
\citep{Burrows05, Wheeler08, Delaney10}.  The combination of LE and
SNR data paint a coherent picture of the SN, and further connect
observations to the explosion.

Our detection of blueshifted H$\alpha$ and \ion{He}{1} emission from
the direction of the NW outflow (LE2521) is the first direct
connection between significantly higher kinetic energy per unit mass
in the SN explosion in one direction and outflows observed in the SNR.
Additionally, the NW outflow that seems to be associated with LE2521
is in roughly the opposite direction of the apparent motion of a
compact X-ray source, which is presumed to be the resulting neutron
star (NS).  From the position of the NS relative to the kinematic
center of the SNR and the age of the SNR,
\citet{Fesen06_expansionasym_age} determined that the object is moving
at 350~\kms\ in the plane of the sky with a position angle of
$169\arcdeg \pm 8.4\arcdeg$.  A simple explanation of both the
high-velocity ejecta in the direction of LE2521 and the NS kick
direction and velocity is an inherently asymmetric explosion.

Although this alignment could be coincidental, it may provide a key clue
to understanding core-collapse SN explosions.  In particular, these
observations indicate that the explosion mechanism may be directly
connected to the observed SN --- a nontrivial statement considering
the size scales between the core and outer envelope of the star.

One model to give the NS a kick is the ejecta-driven
mechanism, in which asymmetry in density before the collapse leads to
an asymmetric explosion. This gives the protoneutron star a kick in
the opposite direction than the ejecta, as in the rocket effect
\citep{Burrows96,Fryer04}. However, if neutrino-driven kicks help
drive the supernova explosion, then the strongest ejecta motion can be
in the direction of the NS, and opposite the direction of
the neutrinos \citep{Fryer06,Socrates05}.

Numerical models have shown that large asymmetric instabilities can occur
near the core during collapse \citep[e.g.,][]{Blondin03, Burrows07,
Marek09}.  These instabilities can be imprinted on both the SN ejecta
and NS by aligning high-velocity ejecta along the same axis as the NS
kick.  3D models have only recently been explored, and
the models have not been extended to examine the effects of this
mechanism on the composition, density, or velocity structure of the
ejecta.  Nonetheless, these mechanisms are an example of a possible
way to directly connect the NS kick with an asymmetric velocity in the
ejecta.


To determine if the alignment of the NS kick and higher-velocity
ejecta can be physically explained, we perform a simple calculation
using the ejecta-driven model.  If we assume that in the direction of
the NS's motion the ejecta velocity is the same as in the
lower-velocity directions we have measured, and that the momentum of
the NS is equal to the excess momentum of the ejecta in the opposite
direction, we can measure the mass of ejecta at higher velocity.  The
velocity difference is \about 4000~\kms\ for the different LoS.  We
also assume that $1 {\rm M}_{\sun} \le M_{\rm NS} \le 2 {\rm
M}_{\sun}$.  Then we have
\begin{equation}
  M_{\rm HV\,ej} \approx 0.09 \left ( \frac{v_{\rm NS}}{350 {\rm \,km\,s}^{-1}} \right ) \left ( \frac{4000 {\rm \,km\,s}^{-1}}{\Delta v_{\rm HV\,ej}} \right ) M_{\rm NS}.
\end{equation}
\noindent
For our assumed NS mass, we find that the amount of material moving at
the higher velocity would be $0.09 \lta M_{\rm HV\,ej} \lta 0.18\,
{\rm M}_{\sun}$.  This mass estimate assumes that the radial velocity
of the NS is zero.  Although the radial velocity could be quite large,
it is likely of order the transverse velocity, and will not
affect our mass estimate by more than a factor of a few. Similarly, we
have assumed that the velocity difference in the ejecta is the total
velocity difference, but there could be a significant component
perpendicular to our line of sight.

\citet{Willingale03} determined that the Cas~A SNR had a total mass of
2.2~M$_{\sun}$ with \about 0.4~M$_{\sun}$ of fast-moving ejecta having
an initial velocity of 15,000~\kms, consistent with our measured
velocity but slightly higher than our estimated mass of high-velocity
material.  \citet{Willingale03} claim that \about 90\% of the kinetic
energy of the SN was at the highest velocities, which seems
inconsistent with SN explosion models.  If the Cas~A SN explosion was
similar to these models, which expect most of the kinetic energy to be
at lower velocities (e.g., $\sim 6000$~\kms), then the lower mass
estimate for the high-velocity material seems reasonable.
This implies a kinetic energy of \about $2 \times
10^{50}$~erg located in high-velocity material. The estimate
of the kinetic energy of Cas~A is \about (2--3)$ \times
10^{51}$~erg \citep{Laming03} for an ejected mass of \about
$2 {\rm M}_{\sun}$. These values are very similar to those derived 
for SN~2006aj, the SN~Ic associated with the X-ray flash GRB~060218
\citep[e.g.,][]{Pian06} and which \citet{Mazzali06} suggested to have
been the result of a magnetar event. The analogy with the properties
of Cas~A and the orientation of the motion of the NS suggests that the
Cas~A SN may have produced a magnetar. This would justify the
higher-than average kinetic energy of Cas~A.

A further constraint for explosion models is the different velocities
of H, He, and Ca for LE2521 relative to SNe~1993J and 2003bg.
SNe~1993J and 2003bg have similar velocities for the \ion{Ca}{2} NIR
triplet, and both are consistent with that of LE2521.  The H$\alpha$
velocity of SN~2003bg is significantly larger than that of SN~1993J,
and is consistent with that of LE2521.  However, SNe~1993J and 2003bg
have similar velocities for \ion{He}{1}, with SN~2003bg having a
slightly larger velocity.  Conversely, LE2521 has a much larger
\ion{He}{1} velocity than that of either comparison object.

There are several ways to explain the velocity structure.  First,
since all spectra appear to have similar velocities for \ion{Ca}{2},
it is likely that the feature is forming in a more symmetric region of
the ejecta.  The small differences between the LEs and the relatively
low velocities of the feature suggest that the Ca-emitting region is
more central than that of H or He, and that the central region of the
ejecta is more symmetric than the outer layers.

There are several explanations for the \ion{He}{1} velocity being
higher than that of SN~2003bg, which had the same H$\alpha$ velocity
of LE2521.  In Cas~A the H layer may have been extremely thin (thinner
than in either SN~1993J or SN~2003bg), causing the H and He velocities
to be coincident.  Alternatively, the Cas~A He layer may be more mixed
into the hydrogen layer than in SN~2003bg, causing \ion{He}{1} to
have a velocity similar to that of H$\alpha$.  Finally, the $^{56}$Ni
distribution may have been different in the two objects, causing a
different ionization structure in the outer ejecta.

\section{Conclusions}
\label{sec:conclusions}

We have obtained optical spectra of LEs from three different
perspectives of the Cas~A SN, effectively probing different regions of
the SN photosphere --- the first time that this technique has been
applied to a SN. The spectra are very similar to each other and are
all similar to the prototypical SN~IIb~1993J.  After accounting for
the window function determined by the combination of dust inclination
and slit orientation, we are able to precisely compare Cas~A to other
SNe as well as compare the LEs to each other.  From these comparisons,
two of the three directions have spectra which are indistinguishable
from that of SN~1993J; however, one direction has \ion{He}{1} and
H$\alpha$ P-Cygni features that are significantly blueshifted (\about
4000~\kms) relative to SN~1993J and the other two directions,
indicating a higher ejecta velocity from Cas~A \textit{in that one
  direction}.  This is direct and independent evidence of an
asymmetric explosion.

The spectrum for the discrepant LE has an H$\alpha$ line profile
consistent with that of the high-luminosity SN~IIb~2003bg, but its
\ion{He}{1} $\lambda 5876$ line profile had an even higher velocity
than that of SN~2003bg.  This may indicate that Cas~A had a very thin
hydrogen layer, significant ejecta mixing, or different ionization
structure in this direction.  All LE spectra have \ion{Ca}{2} NIR
triplet line profiles consistent with each other as well as with those
of SNe~1993J and 2003bg.  This suggests that the emitting region of
the Ca is distributed more spherically than that of the H or
He-emitting regions.

Even though there seems to be a ``jet-like'' structure in the NE
corner and a counterjet in the SW corner \citep{Hwang04,
  Fesen06_expansionasym_age}, recent optical and X-ray data from the
Cas~A SNR indicate that the dominant Cas~A SN outflow is in the SE at
a position angle of \about 115\arcdeg, slightly tilted toward the
observer, and its counterpart approximately on the opposite side
\citep{Burrows05, Wheeler08, Delaney10}.  Our detection of a blueshift
looking into the counter outflow in the NW corner is the first direct,
unambiguous, and independent confirmation of this outflow.  It is also
in excellent agreement with the apparent motion of the compact object,
which moves at a position angle of $169\arcdeg \pm 8.4\arcdeg$
\citep{Tananbaum99, Fesen06_CO} away from the center of the SNR.

Finally, we note that the existing surveys for LE features in this
portion of the Galactic plane are far from complete, and that
additional LEs are very likely to be discovered, providing additional
perspectives of the SN in three dimensions.  The inventory of such
features will further illuminate the degree of asymmetry of the SN,
but will also serve the purpose of testing the degree of coherence of
spectra from similar perspectives.  Since the spectrum at a given 
dust-concentration location is the result of integration over an entire
hemisphere of SN photosphere, spectral differences are expected to
vary slowly with changes in the perspective angle.

\acknowledgements

We thank A.\ Becker, A.\ Clocchiatti, A.\ Garg, M.\ Wood-Vasey, and
the referee for useful comments that helped improve the manuscript.
We are grateful to B.\ Jannuzi and H. Schweiker for taking images with
the KPNO 4~m telescope prior to the design of the Keck slit masks, to
J. M. Silverman for assistance with the Keck observations, and to T.\
Delaney for giving us access to her manuscript and 3D data for the
structure of Cas A before publication.  A.R. thanks the Goldberg
Fellowship Program for its support.  D.W. acknowledges support from
the Natural Sciences and Engineering Research Council of Canada
(NSERC).  C.B. acknowledges the Benoziyo Center for Astrophysics for
support at the Weizmann Institute of Science.  D.M. is supported by
grants from FONDAP CFA 15010003, BASAL CATA PFB-06, and MIDEPLAN MWM
P07-021-F. R.P.K is grateful for the support of NSF grant
AST-0907903. A.V.F. is grateful for the support of NSF grant
AST-0908886, the TABASGO Foundation, and NASA grant GO-11114 from the
Space Telescope Science Institute, which is operated by AURA, Inc.,
under NASA contract NAS 5-26555.

Some of the observations were obtained with the Apache Point Observatory
3.5~m telescope, which is owned and operated by the Astrophysical
Research Consortium.  Part of the data presented herein were obtained
at the W. M. Keck Observatory, which is operated as a scientific
partnership among the California Institute of Technology, the
University of California, and NASA; the observatory was made possible
by the generous financial support of the W. M. Keck Foundation.  The
authors wish to recognize and acknowledge the very significant
cultural role and reverence that the summit of Mauna Kea has always
had within the indigenous Hawaiian community; we are most fortunate to
have the opportunity to conduct observations from this mountain.


\bibliographystyle{fapj}
\bibliography{ms}


\end{document}